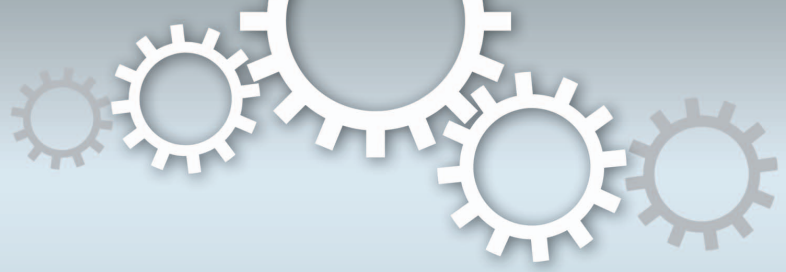

# Observation of inhibited electron-ion coupling in strongly heated graphite

T. G. White[1], J. Vorberger[2], C. R. D. Brown[1,3], B. J. B. Crowley[1,3], P. Davis[4], S. H. Glenzer[4], J. W. O. Harris[3], D. C. Hochhaus[5], S. Le Pape[4], T. Ma[4], C. D. Murphy[1], P. Neumayer[5], L. K. Pattison[3], S. Richardson[3], D. O. Gericke[2] & G. Gregori[1]

[1]University of Oxford, Clarendon Laboratory, Parks Road, Oxford, OX1 3PU, UK, [2]Centre for Fusion, Space and Astrophysics, Department of Physics, University of Warwick, Coventry, CV4 7AL, UK, [3]AWE, Aldermaston, Reading, Berkshire, RG7 4PR, UK, [4]Lawrence Livermore National Laboratory, PO Box 808, Livermore, CA 94551, USA, [5]Extreme Matter Institute, GSI Helmholtzzentrum für Schwerionenforschung, Planckstr. 1, 64291 Darmstadt, Germany.

Creating non-equilibrium states of matter with highly unequal electron and lattice temperatures ($T_{ele} \neq T_{ion}$) allows unsurpassed insight into the dynamic coupling between electrons and ions through time-resolved energy relaxation measurements. Recent studies on low-temperature laser-heated graphite suggest a complex energy exchange when compared to other materials. To avoid problems related to surface preparation, crystal quality and poor understanding of the energy deposition and transport mechanisms, we apply a different energy deposition mechanism, via laser-accelerated protons, to isochorically and non-radiatively heat macroscopic graphite samples up to temperatures close to the melting threshold. Using time-resolved x ray diffraction, we show clear evidence of a very small electron-ion energy transfer, yielding approximately three times longer relaxation times than previously reported. This is indicative of the existence of an energy transfer bottleneck in non-equilibrium warm dense matter.

G raphite is a highly anisotropic semimetal and has proven to be a material with unique electronic[1] and phonon properties[2]. It has the highest electron mobility and thermal conductivity of any bulk material, and both features also suggest unique behaviour in the electron-ion coupling. A large body of work exists studying this electron-ion energy transfer in both graphite[2–9] and graphene[10–12]. However, these measured values strongly disagree with current theoretical models[3]. Since nearly all studies used an optical pump to create a non-thermal distribution of the electrons within the first few tens of nanometers of the sample surface, inhomogeneities in the heating and direct coupling to optical phonons have been suggested as possible sources of uncertainty. While the overall energy transfer between electrons and ions is slow, a strong and almost instantaneous interaction between the excited electrons and high energy ($\sim 200$ meV) coupled optical phonons (COPs) was found[2]. The different channels of energy transfer in carbon act as a complicating factor when modelling electron-ion temperature relaxation within a simple two-temperature model. Additionally, it has been suggested that the electron-ion coupling is further reduced at high electron temperatures[7,13].

Similarly, much longer relaxation times than predicted were also found in extreme states of matter consisting of highly coupled ionic liquids immersed in partially degenerate electrons[14–16]. Such warm dense matter states are highly relevant to understand the evolution and structure of large planets and burned out stars[17–21]. In particular, the phase diagram of carbon is still largely unknown for both the highly exited solid[22] and fluid phases[23]. Here we discuss the results of an experiment where a graphite sample has been heated to very high temperatures by proton beams, but still keeping the lattice structure intact. This allows us to investigate electron-phonon coupling in the high-temperature limit, that is, near melting. Energy deposition by protons also avoids many issues related to laser heating of solids, *e.g.*, direct coupling of the laser light to phonon modes and inhomogeneous heating.

## Results

The experiment has been carried out at the Titan laser at the Jupiter Laser Facility, Lawrence Livermore National Laboratory (USA). Fig. 1 shows a schematic of the experiment. Detailed descriptions of the laser beams, diagnostics and sample setup are given in the methods section. We employed proton beams with energies of up to 10 MeV produced by high-intensity short pulse irradiation of a 10 $\mu$m thick Al foil[24] (see Fig. 2A and Fig. 2B). The protons emerge in a 60° cone from the rear surface of the foil, which is placed 1 mm from the highly orientated pyrolytic graphite (HOPG) sample. This leads to heating within a $\sim 1$ mm diameter region after a time of flight on the order of $t \approx 100$ ps. A 30 $\mu$m plastic layer, placed between the proton beam and the HOPG target, was used to stop electrons, heavier ions and low energy protons, ensuring uniform and predictable heating of the sample. Additionally, the plastic layer acts as a tamper, effectively preventing any thermal expansion on the timescales probed in the experiment. The protons provide isochoric heating throughout the 50 $\mu$m thick HOPG. The heating due to the protons was calculated from measurements of the proton spectrum (see Fig. 2) taken on shots where the HOPG sample was absent and scaled with laser intensity for shots where the sample was present[24]. The measured spectrum was used in a proton ray tracing simulation, in which the energy deposition was modelled using the known stopping power of protons in graphite[25]. Over 99% of the proton energy is deposited in the electron subsystem and the electron temperature accordingly rises to $T_{ele} \approx 17,000$ K (1.5 eV) on a picosecond timescale (see Fig. 2C). Within the electron subsystem, thermalization occurs on a femtosecond timescale. Subsequently, the electron-phonon energy exchange rate determines the temperature evolution of the electron and phonon subsystems.

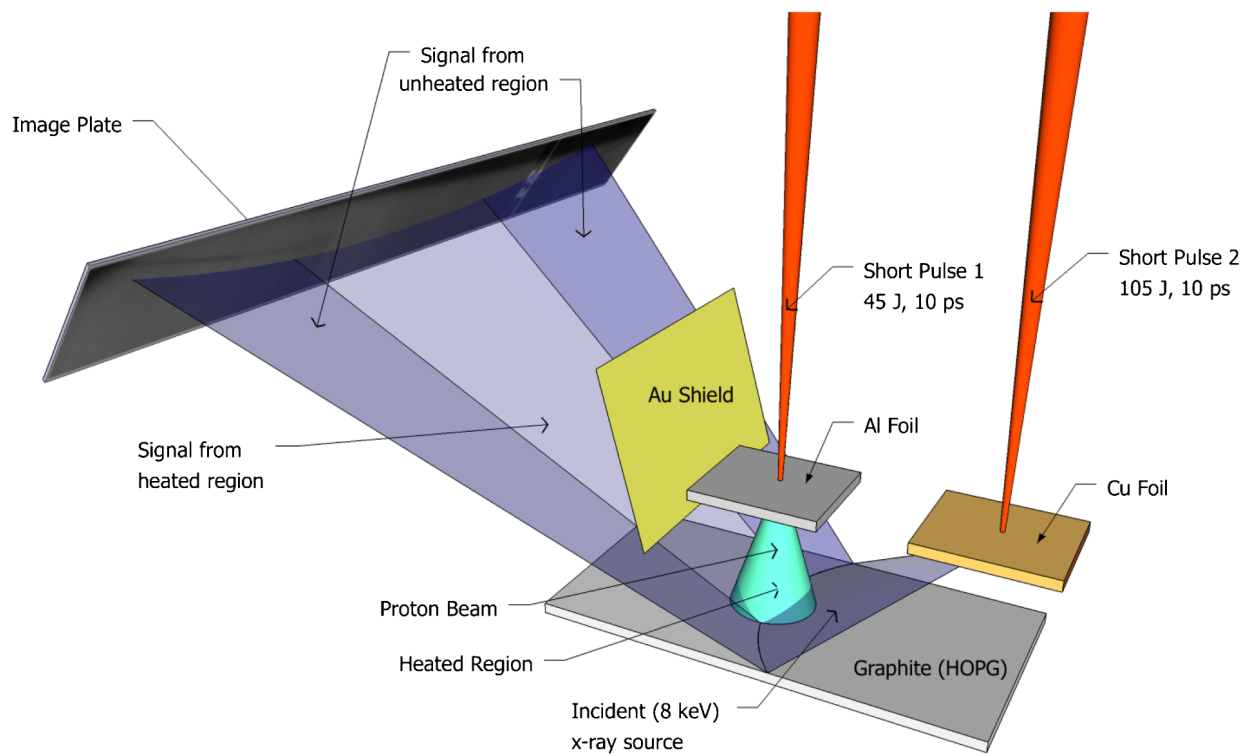

**Figure 1 | A schematic of the experimental setup that was used on the Titan Laser at the Jupiter Laser Facility, Lawrence Livermore National Laboratory (USA).** Additional details on the experiment are given in the methods section.

X rays for probing the sample were generated by a second laser pulse, delayed by $t = 225$ ps with respect to the first pulse, which irradiates a copper foil and, thus, acts as a source of Cu-K$\alpha$ line radiation of 8 keV. Simultaneously, the Bragg scattering of these x rays was measured and the reduction in scattering intensity from the heated regions of the sample was observed. This reduction in intensity is related to increased thermal vibrations of the atoms around their equilibrium lattice sites[26]. Fig. 3A shows the decrease in the intensity of x rays scattered across the surface of the sample. A heated region of approximately 1 mm in width is apparent where the intensity drops to ~30% compared with the scattering from the ambient region. With knowledge of the interatomic potential and the partial phonon density of states (PDOS), the magnitude of thermal vibrations can be connected to the lattice temperature. More specifically, we relate the vibrations to the ratio of lattice and Debye temperatures, $T_{ion}/\Theta_D$, where the Debye temperature characterises the lattice strength through the highest frequency mode in the PDOS. $\Theta_D = 800$ K was extracted from density functional theory calculations and shown to have little variation with electron temperatures in the range considered in our experiment (details of these calculations are given in the supplementary information). Taking into account the effects of absorption and extinction within the sample, the analysis (see Fig. 3B) shows that a drop of 70% in the Bragg intensity is consistent with a lattice temperature of around 3,500 K.

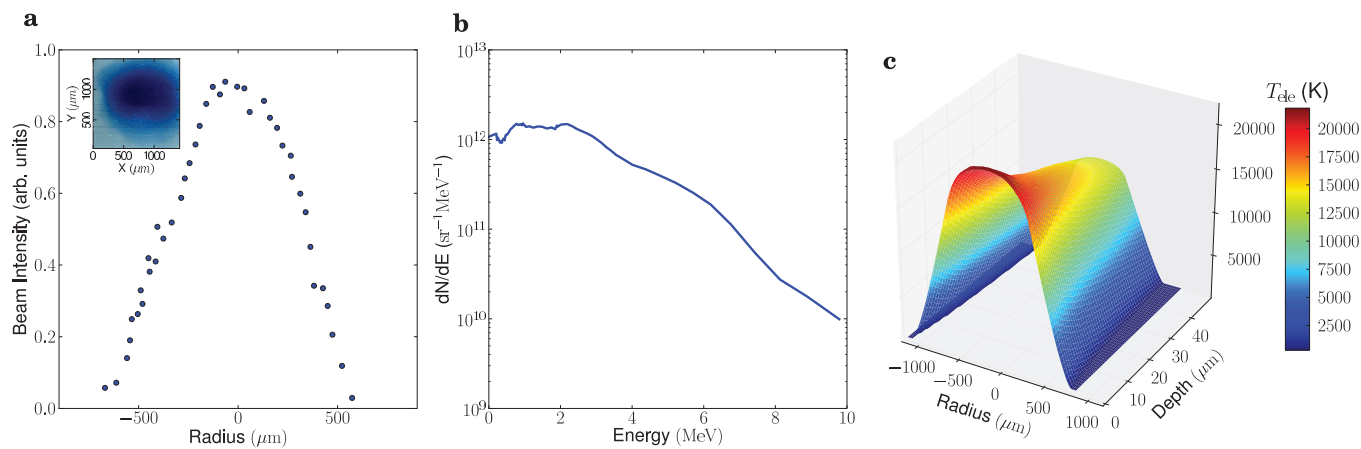

**Figure 2 | Measurement of the proton beam and corresponding electron temperature at t=100 ps.** (a) Measured spatial distribution of the proton intensity showing a ~1 mm width at the position of the sample surface. The inset shows the radiochromic film data used to monitor the proton beam. (b) Proton energy spectrum taken from the centre of the proton beam and plotted up to the maximum energy measured. (c) Calculated electron temperature in the HOPG sample. A ray tracing simulation is used to follow the proton trajectories through the sample. The simulation uses the measured proton distributions in space and energy shown in panels (a) and (b). From the energy deposited by the proton beam and the specific heat of a given equation of state model, the electron temperature throughout the sample can be inferred. Electron heat capacities from density functional simulations are used to obtain the temperature distribution shown.

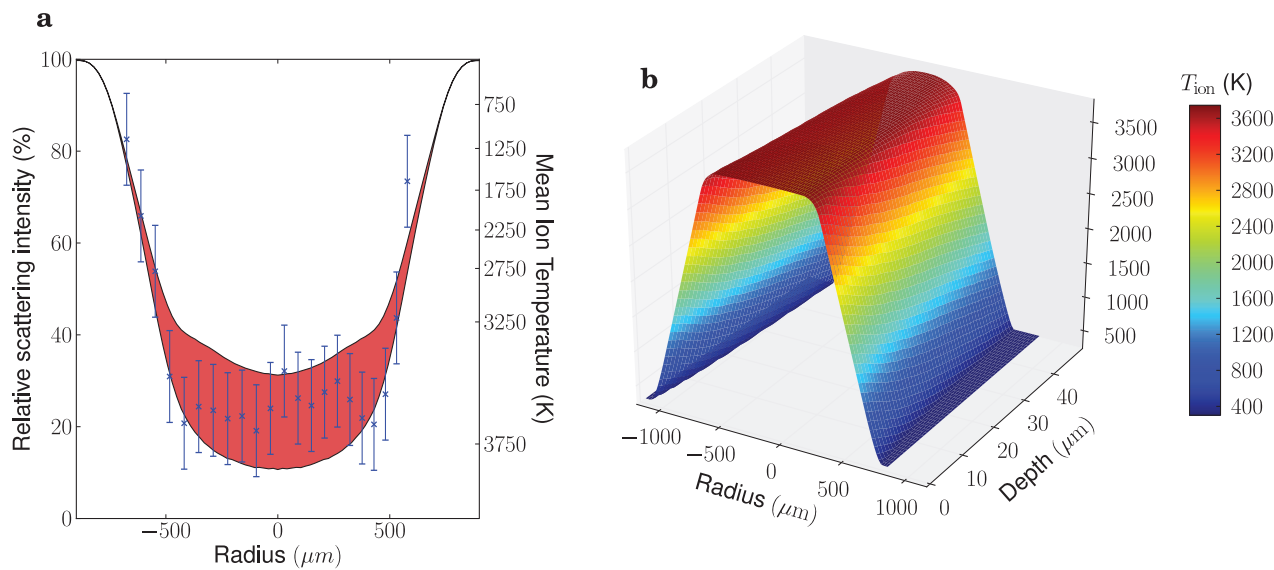

**Figure 3 | Intensity of scattered radiation and corresponding lattice temperature at t=225 ps.** (a) The measured Bragg peak intensity of the scattered radiation from across the sample surface. The values are plotted relative to the Bragg intensity without heating. This ratio is given by $I/I_0 = e^{-2W}$, where $2W = k^2 \langle u_\perp^2 \rangle$ is the Debye-Waller factor. Here, k=3.73 Å$^{-1}$ is the scattering wave number with its direction parallel to the c axis of the graphite lattice and $\langle u_\perp^2 \rangle$ is the out-of-plane rms deviation from the equilibrium ion positions. Since $\langle u_\perp^2 \rangle$ is a function of $T_{ion}/\Theta_D$ (see supplementary information), we can calculate the ion temperature from the measured decrease in the Bragg scattering intensity, as reported on the right axis of the plot. The shaded region corresponds to ion temperatures obtained by solving the two-temperature model (1) and (2) with an electron-phonon coupling constant of g=0.45−0.8×10$^{16}$ W K$^{-1}$ m$^{-3}$. (b) Calculated spatial distribution of the lattice temperature within the graphite sample based on the measured decrease in Bragg scattering (best agreement with experimental data is obtained by assuming g=0.6×10$^{16}$ W K m.).

The average electron-phonon coupling during the experiment can now be obtained from the two-temperature model using the initial electron temperature and the lattice temperature at $t = 225$ ps as boundary condition. The evolution of the temperatures of the two subsystems can be described by

$$C_{ele}\frac{\partial T_{ele}}{\partial t} = \nabla \cdot (K_{ele} \nabla T_{ele}) - g(T_{ele} - T_{ion}) + S_{proton}(\mathbf{r},t), \quad (1)$$

$$C_{ion}\frac{\partial T_{ion}}{\partial t} = \nabla \cdot (K_{ion} \nabla T_{ion}) - g(T_{ele} - T_{ion}), \quad (2)$$

where $C_{ele/ion}$ are the specific heat capacities of the components, $K_{ele/ion}$ the thermal conductivities (which are negligible for the time scale of this experiment), and g is the electron-ion coupling coefficient. An equation of state model is required to relate the energy density to the temperature of each subsystem and, hence, calculate the heat capacities $C_{ele}$ and $C_{ion}$. In this work, we use two different equation of state models: PROPACEOS 4.2, which is based on a chemical model[27] and includes many adjustments based on experiments, and ab initio density functional theory for the electrons coupled to classical molecular dynamics for the ions (DFT-MD). The proton heating term, $S_{proton}(\mathbf{r}, t) \approx S_{proton}(\mathbf{r}, t = 100$ ps$)$, was calculated as described above. This provides the initial condition, and the energy relaxation then evolves according to Eqs. (1) and (2). The coupling coefficient itself is difficult to model from first principles as it depends on the interaction dynamics of the electron and phonon subsystems. However, it allows the complexity of the physical system to be contained within a single parameter and, by scaling g to fit the observed lattice temperature at t=225 ps, we are able to infer its value from the experiment. Additional details on this procedure can be found in the supplementary information.

The results from the fit for g are shown in Fig. 4. The energy deposited by the proton beam heated the sample close to melting, but the lattice remained intact as we still see Bragg scattering. The extracted electron-phonon coupling coefficient is found to be in the range of g = 0.54 – 0.66 × 10$^{16}$ W K$^{-1}$ m$^{-3}$. While we have assumed a constant coupling coefficient, it is expected that g varies with the electron and ion temperatures. Hence, our results can be interpreted of as an average over the material conditions during the first 225 ps of equilibration. The 5% error margin in the intensity of scattered light results from the variance in intensity present in unheated shots. It is also evident from Fig. 4 that only calculations based on the equation of state model from DFTMD are able to match data at t = 225 ps. Indeed, it has previously been shown that chemical models for the equation of state are not sufficiently accurate for describing warm dense graphite, principally due to the strong underestimation of the latent heat[23].

## Discussion

In order to compare with previously published data in the lower temperature range, we transfer our measurements to the dimensionless parameter $\lambda$ that is related to the coupling coefficient by $g = 3\lambda\hbar C_{ele}\langle\omega^2\rangle/\pi k_B T_{ele}$, where $\langle\omega^2\rangle \approx \Theta_D^2/2$ is the second moment of the PDOS[28]. Converting our values for the coupling coefficient gives $\lambda \sim 0.06$. This is considerably lower than the value for typical metals, however previous studies on weakly excited carbon have also found low coupling in room temperature graphite[2,3] ($\lambda < 0.2$), and graphene[10] ($\lambda = 0.05 - 0.1$). These results have been attributed to a small density of states at the Fermi energy in semimetals, an argument that becomes questionable for the high electron temperatures present in our experiment.

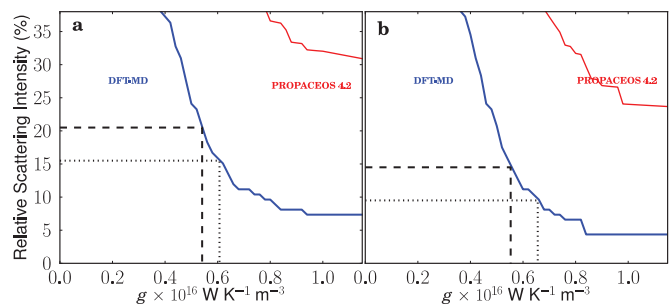

**Figure 4 | Relative Bragg scattering intensity at the centre of the graphite sample at $t=225$ ps plotted against the electron-phonon coupling coefficient.** The calculations are performed by solving equations (1) and (2) with two different models for the equation of state and the heat capacities: PROPACEOS 4.2 (red line) and DFT-MD (blue line). The experimentally measured scattering intensity is shown by the dotted and dashed lines, where the error margin of 5% was estimated from the variation in intensity across unheated samples. (a) The energy of the laser producing the proton beam is $E_{laser}=39.3$ J. (b) $E_{laser}=46.5$ J incident on the Al foil.

There is still much debate over the coupling rate in semimetal materials such as graphite. Recently published work focuses on the coupling between the electrons and COPs at frequencies of $\omega_{COP} \approx 200$ meV. This pathway is believed to allow for a fast energy flow on timescales less than 100 fs, although the decay of these excited phonons occurs over significantly longer times. In experiments using laser heating, the possibility that these optical phonons are excited directly by the pump laser cannot be dismissed[2]. By applying heating through an energetic proton beam, we eliminate direct COP generation by the heating source. Additionally, we probe lower energy phonons with a component of motion out-of-plane through the measurement of the 002 Bragg peak intensity. Therefore we expect no contribution to intensity reduction from COPs[8]. While our method does not directly discriminate between direct coupling and coupling via COPs, the low relaxation rates observed here suggest either saturation of the COP pathway or lower coupling via more direct pathways.

Due to the large electron temperature ($T_{ele} \approx 17,000$ K) achieved, it may be interesting to compare the result found here with data obtained in other high temperature systems in the warm dense matter region, which might be particularly insightful as the coupling coefficient is expected to vary strongly with electron temperature[13]. The experimental difficulties in creating and measuring temperatures within highly non-equilibrium dense systems mean that only a few data points exist; none to date for graphite or carbon. However, experiments performed on warm dense silicon and aluminium samples also suggests a very low coupling parameter[14,15] of the order of $10^{16}$ W K$^{-1}$ m$^{-3}$. To explain these experiments, it has been proposed that the effective electron-ion interaction is suppressed by virtue of a dynamical coupling of electron and ion modes[29–31]. These coupled ion modes mainly occur at frequencies $\omega \ll k_e c_s \simeq 200$ meV, where $k_e$ is the inverse of the Thomas-Fermi screening length and $c_s \sim (k_B T_{ele}/M)^{1/2}$ is the isothermal sound speed ($M$ is the carbon ion mass) suggesting that low frequency excitations have an important role in determining the pathway and the bottleneck in the energy relaxation. To interpret our experimental results, we have to assume that frequencies $\omega \ll \omega_{COP}$ play the dominant role in the electron-phonon energy exchange. Such processes may thus be more general than previously believed and, as indicated by previous experiments, occur in any strongly driven two-temperature system. Moreover, we can expect similar changes in other transport properties such as the optical conductivity in dense plasmas, having wide implications from laser processing of materials and inertial confinement fusion[16] to the cooling rates of white dwarf stars and neutron stars[17,21]. The work presented here will be generally applicable to other highly coupled systems as a method to probe the potential landscape surrounding the ions as long as the lattice remains intact. As such it is perfectly suited to studying temperature relaxation in many highly non-equilibrium solids and warm dense matter systems[16].

## Methods

The experiment was performed at the Titan laser at the Jupiter Laser Facility, Lawrence Livermore National Laboratory (USA). The 150 J, 10 ps Titan beam at the fundamental wavelength $\lambda_0=1054$ nm was split into two arms in a pump-probe configuration. To heat the graphite sample, 30% of the energy was delivered onto a thin aluminum foil with a 10 μm focal spot. The accelerated electrons create an electric field which drives an intense proton beam normal to the rear surface of the aluminum foil. The remaining energy was focused 225 ps later with a 10 μm focal spot onto a copper disk to generate Cu-Kα probe x-rays. The image plate detector was placed 300 mm away from the graphite sample at 27.4° to detect the Bragg diffracted light. The magnification of the setup was 138. The proton beam forms a spot of diameter ~1.0 mm on the sample. The sample consisted of 50 μm thick HOPG covered with 30 μm of polypropylene.

### Acknowledgements
We would like to thank all the staff at the Jupiter Laser Facility for their support. This research leading to these results was supported by EPSRC grants EP/G007187/1, EP/D062837, and EP/I014888/1. Partial support from AWE and from the Helmholtz Alliance EMMI is acknowledged. This work also received funding from LDRD grant No. 11-ERI-050 and was performed under the auspices of the U.S. Department of Energy by Lawrence Livermore National Laboratory under Contract No. DE-AC52-07NA27344. The authors would like to thanks Prof. Don Lamb and the FLASH team at the University of Chicago for discussions on this topic.

### Author contribution
G.G. conceived this project and the experimental work was performed by P.D., D.C.H, S.L.P., T.M., P.N. and T.G.W. The analysis was completed by T.G.W. and C.D.M. while the DFT simulations performed by T.G.W. and J.V. The paper was written by T.G.W., G.G., D.O.G., J.V. and B.J.B.C. Additional theoretical and experimental support was provided by C.R.D.B., S.H.G., J.W.O.H., L.K.P. and S.R.

### Additional information
**Supplementary information** accompanies this paper at http://www.nature.com/scientificreports

**Competing financial interests:** The authors declare no competing financial interests.

**How to cite this article:** White, T.G. *et al.* Observation of inhibited electron-ion coupling in strongly heated graphite. *Sci. Rep.* **2**, 889; DOI:10.1038/srep00889 (2012).